\documentclass[10pt,thmsa]{article}
\usepackage{graphics}
\input{tcilatex}

\begin{document}

\begin{center}
{\Large Is there the color-flavor locking in the instanton induced
quark-antiquark pairing in QCD vacuum? }

\vskip 0.4cm
%
\textbf{Hoang Ba Thang}\textit{, }

\textit{Institute of Materials Science and Institute of Physics - NCST of Vietnam}
\end{center}

\begin{quotation}
\vskip 0.5cm
%
\noindent {\small By means of the functional integral method we show that in
the case of the quark-antiquark pairing at zero temperature and zero
chemical potential (in the vacuum) the singlet pairing is more preferable
than that with the color-flavor locking (CFL).}
\end{quotation}

\section{Introduction}

Recently, the physical properties of the existence of the four-quark
interactions (color superconductivity, the quark-antiquark pairing and the
formation of unusual bound states of quark matter, etc...) were widely
discussed$^{[1-10]}$. In an earlier work $^{[1]}$, Mark Alford, Krishma
Rajagopal and Frank Wilczek have proposed that due to the condensation of
quarks the chiral flavor $SU\left( 3\right) $ symmetries are broken and the $%
LL$ condensate ``locks'' $SU\left( 3\right) _{L}$ rotations to color
rotations while the $RR$ condensate ``locks'' $SU\left( 3\right) _{R}$
rotations to color rotations:

\[
SU\left( 3\right) \otimes SU\left( 3\right) _{L}\otimes SU\left( 3\right)
_{R}\otimes U\left( 1\right) _{B}\rightarrow SU\left( 3\right) _{C+L+R} 
\]
In this work the problem of the color-flavor locking in the quark-antiquark
pairing is considered. We calculate the free energy densities of the QCD
vacuum due to the quark-antiquark pairing in two cases: the singlet pairing
and the pairing with the color-flavor locking. By comparing these values of
the free energy densities we conclude that the singlet pairing is more
preferable.

Let us start from the four-fermion interaction Lagrangian of the four quark
fields

\begin{equation}
L_{\text{int}}\left( x\right) =\frac{1}{2}\overline{\psi }^{A}\left(
x\right) \psi _{B}\left( x\right) U_{AC}^{BD}\overline{\psi }^{C}\left(
x\right) \psi _{D}\left( x\right) ,  \label{1}
\end{equation}
where $\psi _{A}\left( x\right) $ and $\overline{\psi }^{A}\left( x\right) $
are the quark and antiquark fields. $A,B,C,D$ are the sets of indices

\[
A=\left( \alpha ,a,i\right) ,\qquad B=\left( \beta ,b,j\right) ,\qquad
C=\left( \gamma ,c,k\right) ,\qquad D=\left( \delta ,d,l\right) , 
\]
$\alpha ,\beta ,\gamma ,\delta =1,2,3,4$ are the Dirac spinor indices, $%
a,b,c,d=1,2,3...N_{C}$ are the color symmetry indices and $%
i,j,k,l=1,2,3...N_{f}$ are the flavor ones. The coupling constants $%
U_{AC}^{BD}$ are antisymmetric under the permutations of the indices

\[
U_{AC}^{BD}=-U_{AC}^{DB}=-U_{CA}^{BD}=U_{CA}^{DB}. 
\]
The functional integral of the free and interacting massless quark fields are

\begin{equation}
Z_{0}=\int \left[ D\psi \right] \left[ D\overline{\psi }\right] \exp \left\{
-i\int d^{4}x\overline{\psi }^{A}\left( x\right) \left( \stackrel{\wedge }{%
\partial }_{x}\right) _{A}^{B}\psi _{B}\left( x\right) \right\}  \label{2}
\end{equation}
and 
\begin{eqnarray}
Z &=&\int \left[ D\psi \right] \left[ D\overline{\psi }\right] \exp \left\{
-i\int d^{4}x\overline{\psi }^{A}\left( x\right) \left( \stackrel{\wedge }{%
\partial }_{x}\right) _{A}^{B}\psi _{B}\left( x\right) \right\}  \nonumber \\
&&\exp \left\{ \frac{i}{2}\int d^{4}x\overline{\psi }^{A}\left( x\right)
\psi _{B}\left( x\right) U_{AC}^{BD}\overline{\psi }^{C}\left( x\right) \psi
_{D}\left( x\right) \right\} .  \label{3}
\end{eqnarray}

\noindent \noindent Introduce the composite bosonic field $\Phi
_{A}^{B}\left( x\right) $ and the functional integral

\begin{equation}
Z_{0}^{\Phi }=\int \left[ D\Phi \right] \exp \left\{ -\frac{i}{2}\int
d^{4}x\Phi _{A}^{B}\left( x\right) U_{AC}^{BD}\Phi _{D}^{C}\left( x\right)
\right\}  \label{4}
\end{equation}
By means of the Hubbard - Stratonovich transformation, we rewrite $Z$ in the
form of a functional integral over the composite bosonic fields$^{\left[
7,8\right] }$

\begin{equation}
Z=\frac{Z_{0}}{Z_{0}^{\Phi }}\int \left[ D\Phi \right] \exp \left\{ iS_{%
\text{eff}}\left[ \Phi \right] \right\}  \label{5}
\end{equation}
with the effective action $S_{\text{eff}}\left[ \Phi \right] $

\begin{eqnarray}
S_{\text{eff}}\left[ \Phi \right] &=&-\frac{1}{2}\int d^{4}x\Phi
_{A}^{B}\left( x\right) U_{AC}^{BD}\Phi _{D}^{C}\left( x\right) +W\left[
\Delta \right] ,  \label{6} \\
W\left[ \Delta \right] &=&\sum\limits_{n=1}^{\infty }W^{\left( n\right)
}\left[ \Delta \right]  \nonumber
\end{eqnarray}
\begin{eqnarray}
W^{\left( 1\right) } &=&\int dx\Delta _{A}^{B}\left( x\right)
S_{B}^{A}\left( x\right) ,  \nonumber \\
W^{\left( 2\right) } &=&-\frac{1}{2}\int dx_{1}\int dx_{2}\Delta
_{A_{1}}^{B_{1}}\left( x_{1}\right) S_{B_{1}}^{A_{2}}\left(
x_{1}-x_{2}\right) \Delta _{A_{2}}^{B_{2}}\left( x_{2}\right)
S_{B_{2}}^{A_{1}}\left( x_{2}-x_{1}\right) ,  \nonumber \\
W^{\left( 3\right) } &=&\frac{1}{3}\int dx_{1}\int dx_{2}\int dx_{3}\Delta
_{A_{1}}^{B_{1}}\left( x_{1}\right) S_{B_{1}}^{A_{2}}\left(
x_{1}-x_{2}\right) \Delta _{A_{2}}^{B_{2}}\left( x_{2}\right)
S_{B_{2}}^{A_{3}}\left( x_{2}-x_{3}\right)  \nonumber \\
&&\Delta _{A_{3}}^{B_{3}}\left( x_{3}\right) S_{B_{3}}^{A_{1}}\left(
x_{3}-x_{1}\right) ,  \nonumber \\
W^{\left( 4\right) } &=&......  \label{7}
\end{eqnarray}
and $\Delta _{A}^{B}$ play the role of the order parameters of the system of
quarks

\begin{equation}
\Delta _{B}^{A}\left( x\right) =U_{AC}^{BD}\Phi _{D}^{C}\left( x\right) .
\label{8}
\end{equation}
From the effective action (6) we can derive the field equation by means of
the variational principle

\begin{equation}
\Delta _{C}^{D}\left( x\right) =U_{AC}^{BD}\frac{\delta W\left[ \Delta
\right] }{\delta \Delta _{B}^{A}\left( x\right) }=\Delta _{C}^{D}\left(
x\right) G_{B}^{A}\left( x,x\right)  \label{9}
\end{equation}
with the interacting Green's function $G_{B}^{A}\left( x,y\right) $
satisfying the Schwinger-Dyson equations

\begin{equation}
G_{B}^{A}\left( y,x\right) =G_{B}^{A}\left( y-x\right) -\int
dzG_{B}^{C}\left( y-z\right) \Delta _{C}^{D}\left( z\right) G_{D}^{A}\left(
z,x\right) .  \label{10}
\end{equation}
$S_{B}^{A}\left( y-x\right) $ is the Green's function of free quarks.

\section{Equations for the order parameters}

We consider the special class of solution of the equation (9)

\begin{eqnarray}
\Delta _{A}^{B}\left( x\right) &=&\Delta _{A}^{B}=\text{constant},  \nonumber
\\
G_{B}^{A}\left( x,y\right) &=&G_{B}^{A}\left( x-y\right) .  \label{11}
\end{eqnarray}
In this case (10) becomes

\begin{equation}
G_{B}^{A}\left( x-y\right) =S_{B}^{A}\left( x-y\right) -\int
dzS_{B}^{C}\left( x-z\right) \Delta _{C}^{D}G_{D}^{A}\left( z-y\right) .
\label{12}
\end{equation}
Then equation (9) can be written in the form

\begin{equation}
\Delta _{D}^{C}=U_{CA}^{DB}\frac{1}{\left( 2\pi \right) ^{4}}\int d^{4}p%
\widetilde{G}_{B}^{A}\left( p\right) ,  \label{13}
\end{equation}
Denote $\widetilde{G}_{B}^{A}\left( p\right) $the Fourier component of $%
G_{B}^{A}\left( x-x\right) $

\begin{equation}
\left[ \frac{1}{\widetilde{G}\left( p\right) }\right] _{B}^{A}=\left[ \frac{1%
}{\widetilde{S}\left( p\right) }\right] _{B}^{A}+\Delta _{B}^{A}.  \label{14}
\end{equation}
In the case of the singlet pairing the order parameters $\Delta _{B}^{A}$
equal

\begin{equation}
\Delta _{B}^{A}=\delta _{\beta }^{\alpha }\Delta _{bj}^{ai}=\delta _{\beta
}^{\alpha }\delta _{b}^{a}\delta _{j}^{i}\Delta  \label{15}
\end{equation}
with some constant $\Delta .$ It follows that in the limiting case with zero
temperature $\left( T=0\right) $ and zero chemical potential $\left( \mu
=0\right) ,$ the Green's function has the form

\begin{equation}
\left[ \frac{1}{\widetilde{G}\left( p\right) }\right] _{B}^{A}=\left( i%
\widehat{p}+M+\Delta \right) _{\beta }^{\alpha }\delta _{b}^{a}\delta
_{j}^{i}.  \label{16}
\end{equation}
It is clearly that the chiral symmetry is broken due to the existence of $%
\Delta $. In the case of the color-flavor locking the order parameters $%
\Delta _{B}^{A}$ equal

\begin{equation}
\Delta _{B}^{A}=\delta _{\beta }^{\alpha }\left\{ \delta _{b}^{a}\delta
_{j}^{i}\Delta +\delta _{j}^{a}\delta _{b}^{i}\Omega \right\} .  \label{17}
\end{equation}
Then the Green's function has the expression of the form

\begin{equation}
\widetilde{G}_{B}^{A}\left( p\right) =\left( \widehat{A}\right) _{\beta
}^{\alpha }\delta _{b}^{a}\delta _{j}^{i}+\left( \widehat{B}\right) _{\beta
}^{\alpha }\delta _{j}^{a}\delta _{b}^{i}  \label{18}
\end{equation}

\begin{equation}
\widehat{A}=i\widehat{p}X+Y,\qquad \widehat{A}=i\widehat{p}Z+W,  \label{19}
\end{equation}
$X,Y,W$ and $Z$ are scalar functions need to be determined.

\noindent In the systems with the instanton induced interaction between
quarks and antiquarks the coupling constants equal$^{\left[ 4\right] }$

\begin{eqnarray}
U_{AC}^{BD} &=&\left( \delta _{i}^{j}\delta _{k}^{l}-\delta _{k}^{j}\delta
_{i}^{l}\right) \left\{ U_{1}\left[ \left( \delta _{a}^{\beta }\delta
_{\gamma }^{\delta }+\left( \gamma _{5}\right) _{\alpha }^{\beta }\left(
\gamma _{5}\right) _{\gamma }^{\delta }\right) \delta _{a}^{b}\delta
_{c}^{d}+\left( \delta _{a}^{\delta }\delta _{\gamma }^{\beta }+\left(
\gamma _{5}\right) _{\gamma }^{\beta }\left( \gamma _{5}\right) _{\alpha
}^{\delta }\right) \delta _{a}^{d}\delta _{c}^{b}\right] \right.  \nonumber
\\
&&+\left. U_{2}\left[ \left( \sigma _{\mu \nu }\right) _{\alpha }^{\beta
}\left( \sigma _{\mu \nu }\right) _{\gamma }^{\delta }\delta _{a}^{b}\delta
_{c}^{d}+\left( \left( \sigma _{\mu \nu }\right) _{\alpha }^{\delta }\left(
\sigma _{\mu \nu }\right) _{\gamma }^{\beta }\right) \delta _{a}^{d}\delta
_{c}^{b}\right] \right\} .  \label{20}
\end{eqnarray}
In order to determine $X,Y,W$ and $Z$ we insert (17 ), (18 ) and (20 ) into
(13 ). Because the integrals of the terms containing the factor $\widehat{p}$
vanish $X$ and $Z$ give no contribution. For $Y$ and $W$ we have the
expressions

\begin{eqnarray}
Y &=&\frac{1}{2}\left\{ \frac{M+\Delta +\Omega }{p^{2}+\left( M+\Delta
+\Omega \right) ^{2}}+\frac{M+\Delta -\Omega }{p^{2}+\left( M+\Delta -\Omega
\right) ^{2}}\right\}  \nonumber \\
W &=&\frac{1}{2}\left\{ \frac{M+\Delta +\Omega }{p^{2}+\left( M+\Delta
+\Omega \right) ^{2}}-\frac{M+\Delta -\Omega }{p^{2}+\left( M+\Delta -\Omega
\right) ^{2}}\right\} ,  \label{21}
\end{eqnarray}
It follows that $\Delta $ and $\Omega $ are determined by 
\begin{eqnarray}
\Delta &=&\left( N_{C}-1\right) \left( N_{C}U_{1}+U_{2}\right) \frac{1}{%
\left( 2\pi \right) ^{4}}\int d^{4}pY  \nonumber \\
&&+\left[ \left( N_{C}-1\right) U_{1}+U_{2}\right] \frac{1}{\left( 2\pi
\right) ^{4}}\int d^{4}pW,  \label{22} \\
\Omega &=&-U_{2}\frac{1}{\left( 2\pi \right) ^{4}}\int d^{4}pW.  \label{23}
\end{eqnarray}
$N_{C}$ is number of colors. In reference [4] it was shown that the
constants $U_{1}$ and $U_{2}$ equal

\begin{equation}
U_{1}=\frac{4g\left( 2N_{C}-1\right) }{2N_{C}\left( N_{C}^{2}-1\right) }%
,\qquad U_{2}=\frac{g\left( N_{C}-4\right) }{2N_{C}\left( N_{C}^{2}-1\right) 
},  \label{24}
\end{equation}
with some coupling constant $g$. Using this result with $N_{C}=3$ we rewrite
(22) and (23) in the more compact form

\begin{eqnarray}
\Delta +\Omega &=&\frac{g}{24}\frac{1}{16\pi ^{2}}\int p^{2}dp^{2}\left[ 42%
\frac{M+\Delta +\Omega }{p^{2}+\left( M+\Delta +\Omega \right) ^{2}}+22\frac{%
M+\Delta -\Omega }{p^{2}+\left( M+\Delta -\Omega \right) ^{2}}\right] ,\qquad
\label{25} \\
\Delta -\Omega &=&\frac{g}{24}\frac{1}{16\pi ^{2}}\int p^{2}dp^{2}\left[ 44%
\frac{M+\Delta +\Omega }{p^{2}+\left( M+\Delta +\Omega \right) ^{2}}+20\frac{%
M+\Delta -\Omega }{p^{2}+\left( M+\Delta -\Omega \right) ^{2}}\right] .
\label{26}
\end{eqnarray}
If there is no CFL, $\Omega =0$, then the two equations for $\Delta \,$and $%
\Omega $ reduce to one and the same equation for $\Delta $

\begin{equation}
\Delta =\frac{g}{6\pi ^{2}}\int p^{2}dp^{2}\frac{M+\Delta }{p^{2}+\left(
M+\Delta \right) ^{2}}.  \label{27}
\end{equation}
For the convenient we set

\begin{equation}
\Delta _{1}=\Delta +\Omega ,\qquad \Delta _{2}=\Delta -\Omega  \label{28}
\end{equation}
then after taking the integrations over $p^{2}$ we rewrite the equations
(25), (26) and (27) in the form

\begin{eqnarray}
\Delta _{1} &=&\frac{1}{32}\frac{g}{6\pi ^{2}}\left\{ 21\left( M+\Delta
_{1}\right) \left[ 1-\left( M+\Delta _{1}\right) ^{2}\ln \frac{1+\left(
M+\Delta _{1}\right) ^{2}}{\left( M+\Delta _{1}\right) ^{2}}\right] \right. 
\nonumber \\
&&\left. +11\left( M+\Delta _{2}\right) \left[ 1-\left( M+\Delta _{2}\right)
^{2}\ln \frac{1+\left( M+\Delta _{2}\right) ^{2}}{\left( M+\Delta
_{2}\right) ^{2}}\right] \right\} ,  \label{29} \\
\Delta _{1} &=&\frac{1}{32}\frac{g}{6\pi ^{2}}\left\{ 22\left( M+\Delta
_{1}\right) \left[ 1-\left( M+\Delta _{1}\right) ^{2}\ln \frac{1+\left(
M+\Delta _{1}\right) ^{2}}{\left( M+\Delta _{1}\right) ^{2}}\right] \right. 
\nonumber \\
&&\left. +10\left( M+\Delta _{2}\right) \left[ 1-\left( M+\Delta _{2}\right)
^{2}\ln \frac{1+\left( M+\Delta _{2}\right) ^{2}}{\left( M+\Delta
_{2}\right) ^{2}}\right] \right\} ,  \label{30} \\
\Delta &=&\frac{g}{6\pi ^{2}}\left( M+\Delta \right) \left[ 1-\left(
M+\Delta \right) ^{2}\ln \frac{1+\left( M+\Delta \right) ^{2}}{\left(
M+\Delta \right) ^{2}}\right] .  \label{31}
\end{eqnarray}
The values of $\Delta _{1},\Delta _{2}$ and $\Delta $ are plotted in Figures
1, 2 and 3 with certain values of $g$ and $M.$ Note that in the case of
massless quarks, $M=0$, from (29), (30) and (31) it follows that

\begin{equation}
\Delta _{1}=\Delta _{2}=\Delta  \label{32}
\end{equation}
This means that $\Omega $ must equal to zero. Thus we conclude that in the
case of the massless quarks there is no CFL.

\section{The free energy densities}

In order to choose the more preferable one among two pairing mechanisms in
the case $M\neq 0$ we must calculate the free energy density$^{\left[
7,8\right] }$

\begin{eqnarray}
F &=&\frac{1}{\left( 2\pi \right) ^{4}}\int d^{4}p\text{Tr}\left\{ \frac{1}{2%
}\left( \left[ \widehat{\Delta }\widehat{S}\left( p\right) \right] -\left[ 
\widehat{\Delta }\widehat{S}\left( p\right) \right] ^{2}+\left[ \widehat{%
\Delta }\widehat{S}\left( p\right) \right] ^{3}-\left[ \widehat{\Delta }%
\widehat{S}\left( p\right) \right] ^{4}+...\right) \right.  \nonumber \\
&&-\left. \left( \left[ \widehat{\Delta }\widehat{S}\left( p\right) \right] -%
\frac{1}{2}\left[ \widehat{\Delta }\widehat{S}\left( p\right) \right] ^{2}+%
\frac{1}{3}\left[ \widehat{\Delta }\widehat{S}\left( p\right) \right] ^{3}-%
\frac{1}{4}\left[ \widehat{\Delta }\widehat{S}\left( p\right) \right]
^{4}+...\right) \right\}  \nonumber \\
&=&\frac{1}{\left( 2\pi \right) ^{4}}\int d^{4}p\text{Tr}\left[ \widehat{%
\Delta }\widehat{S}\left( p\right) \left\{ \frac{1}{2}\frac{1}{1+\widehat{%
\Delta }\widehat{S}\left( p\right) }-\int\limits_{0}^{1}d\alpha \frac{1}{%
1+\alpha \widehat{\Delta }\widehat{S}\left( p\right) }\right\} \right] .
\label{33}
\end{eqnarray}
If we define a new Green's function $\widetilde{G}\left( p,\alpha \right) $

\begin{equation}
\left[ \frac{1}{\widetilde{G}\left( p,\alpha \right) }\right]
_{B}^{A}=\left[ \frac{1}{\widetilde{S}\left( p\right) }\right]
_{B}^{A}+\alpha \Delta _{B}^{A}  \label{34}
\end{equation}
then we can write the free energy $F$ as

\begin{equation}
F=\frac{1}{\left( 2\pi \right) ^{4}}\int d^{4}p\left[ \widehat{\Delta }%
\left\{ \frac{1}{2}\widetilde{G}\left( p\right) -\int\limits_{0}^{1}d\alpha 
\widetilde{G}\left( p,\alpha \right) \right\} \right] .  \label{35}
\end{equation}
We see here that the free energy is a function of the order parameters and
then with the support of the calculation of $\Delta _{B}^{A}$ in Section II
we can write down the free energy densities of the two cases $F\left( \Delta
\right) $ and $F\left( \Delta _{1},\Delta _{2}\right) $ as follow

\begin{eqnarray}
F\left( \Delta \right) &=&\Delta \left( M+\Delta \right) \left\{ 1-\left(
M+\Delta \right) ^{2}\ln \frac{1+\left( M+\Delta \right) ^{2}}{\left(
M+\Delta \right) ^{2}}\right\}  \nonumber \\
&&-\left\{ \frac{1}{2}\left( M+\Delta \right) ^{2}+\frac{1}{2}\left[
1-\left( M+\Delta \right) ^{4}\right] \ln \left[ 1+\left( M+\Delta \right)
^{2}\right] \right\}  \nonumber \\
&&+\left\{ \frac{1}{2}M^{2}+\frac{1}{2}\left[ 1-M^{4}\right] \ln \left[
1+M^{2}\right] +\frac{1}{2}M^{4}\ln M^{4}\right\} ,  \label{36} \\
F\left( \Delta _{1},\Delta _{2}\right) &=&\frac{2}{3}F\left( \Delta
_{1}\right) +\frac{1}{3}F\left( \Delta _{2}\right)  \label{37}
\end{eqnarray}
The values of $F\left( \Delta _{1},\Delta _{2}\right) $ and $F\left( \Delta
\right) $ with $\Delta _{1},\Delta _{2}$ and $\Delta $ given by (29), (30)
and (31) are plotted in Figures 4 and 5. We can conclude that the value of $%
F\left( \Delta \right) $ is always lower than that of $F\left( \Delta
_{1},\Delta _{2}\right) $. This means that the singlet pairing is more
preferable than that with the color-flavor locking.

\vskip 0.4cm%
{\large \bf Acknowledgements.}%

The author would like to express his deep thanks to professor Nguyen Van
Hieu for suggesting the problem and helpful discussions. The support of the
National Natural Sciences Council is highly appreciated%
\vskip 1cm%
\begin{description}
\item[\textbf{Figure captions}]  \noindent \noindent \noindent 
\vskip 0.5cm%
\end{description}

\begin{enumerate}
\item[Figure1:]  {}\noindent \noindent $\Delta \,$as a function of coupling
constant $g$ (in GeV$^{-2}$) at $M=0.1Mev$

\item[Figure 2:]  $\Delta _{1}\,$as a function of coupling constant $g$ (in
GeV$^{-2}$) at $M=0.1Mev$

\item[Figure 3:]  $\Delta _{2}\,$as a function of coupling constant $g$ (in
GeV$^{-2}$) at $M=0.1Mev$

\item[Figure 4:]  The values of $F\left( \Delta _{1},\Delta _{2}\right) $
depend on $\Delta _{1}$ and $\Delta _{2}$

\item[Figure 5:]  The values of $F\left( \Delta \right) $ as a function of $%
\Delta $
\end{enumerate}
\newpage
\noindent 
{\large \bf References.}%

\begin{itemize}
\item[{\lbrack 1].}]  Mark Alford, Krishma Rajagopal and Frank Wilczek, 
\textit{hep-ph/9804403.}

\item[{\lbrack 2].}]  D. T. Son,\textit{\ Phys. Rev.} \textbf{D59}
(1999)094019,\textit{\ hep-ph/9812287.}

\item[{\lbrack 3].}]  N. V. Hieu and L. T. Tuong, \textit{Commun. Phys.} 
\textbf{8} (1998) 129.

\item[{\lbrack 4].}]  See, for example, G. t'Hoof., \textit{Phy. Rev}. 
\textbf{D14} (1976) 3432; R. Rapp, T. Schafer, E. Shuyak and M. Velkovsky., 
\textit{Phys. Rev. Lett} \textbf{81} (1998)53; N. V. Hieu, H. N. Long, N. H.
Son and N. A. Viet.,\textit{\ hep-ph/0001234}.

\item[{\lbrack 5].}]  Thomas Sch\"{a}fer and Frank Wilczek, \textit{%
hep-ph/9906512}.

\item[{\lbrack 6].}]  Jurgen Bergs and Krishma Rajagopal, \textit{%
hep-ph/9804233.}

\item[{\lbrack 7].}]  N. V. Hieu, \textit{Basics of Fuctional Integral
Technique in Quantum Field Theory of Many-Body Systems, VNUH Pub}., Hanoi,
1999.

\item[{\lbrack 8].}]  Nguyen Van Hieu., Functional Integral Method in the
Theory of Color Superconductivity, Invited talk at the International
Conference at Extreme Energy-Rencontres du Vietnam IV, Hanoi 19-25 July 2000.

\item[{\lbrack 9].}]  N. V. Hieu, N. H. Son, N. V. Thanh and H. B. Thang,%
\textit{\ hep-ph/0001251.}

\item[{\lbrack 10].}]  V. A. Miransky, I. A. Showkovy and L. C. R.
Wijewardhana, \textit{hep-ph/0003327, hep-ph/0009173}.
\end{itemize}

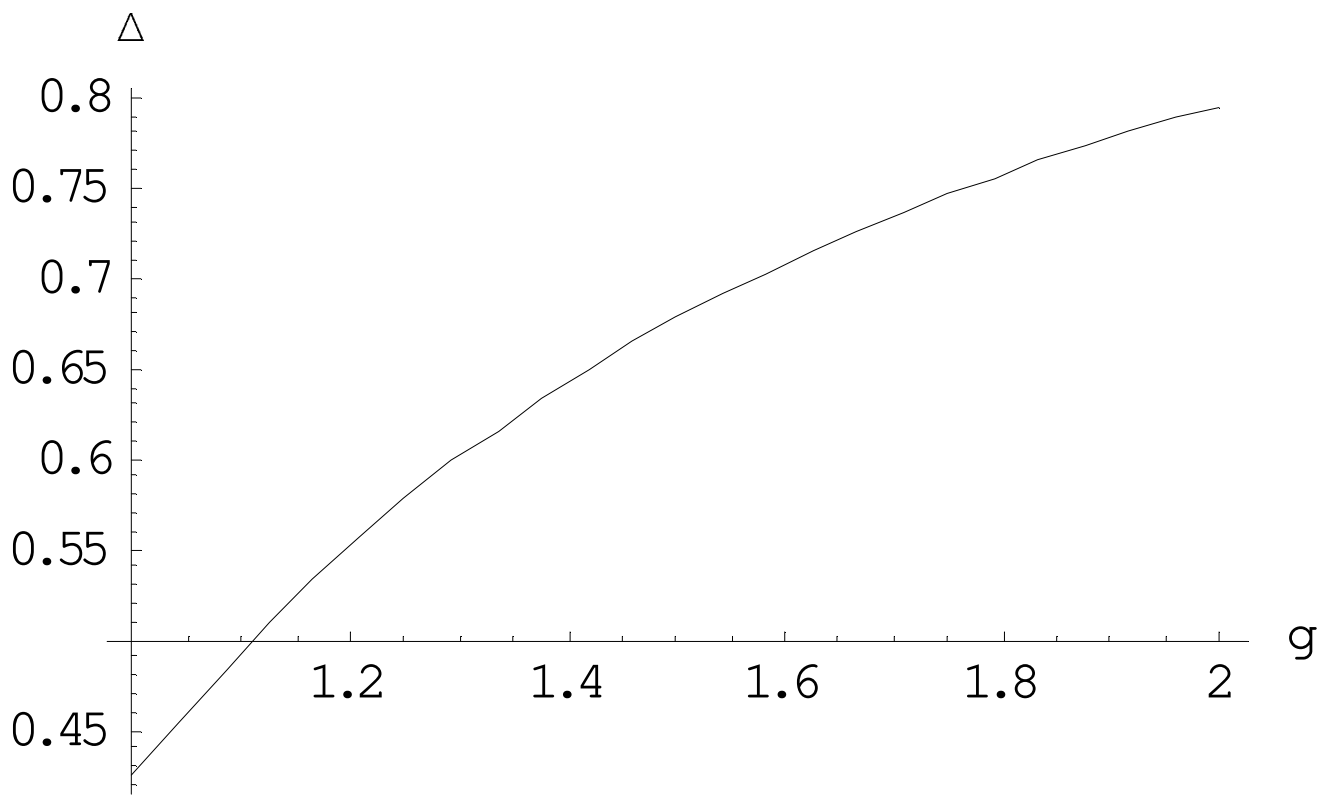
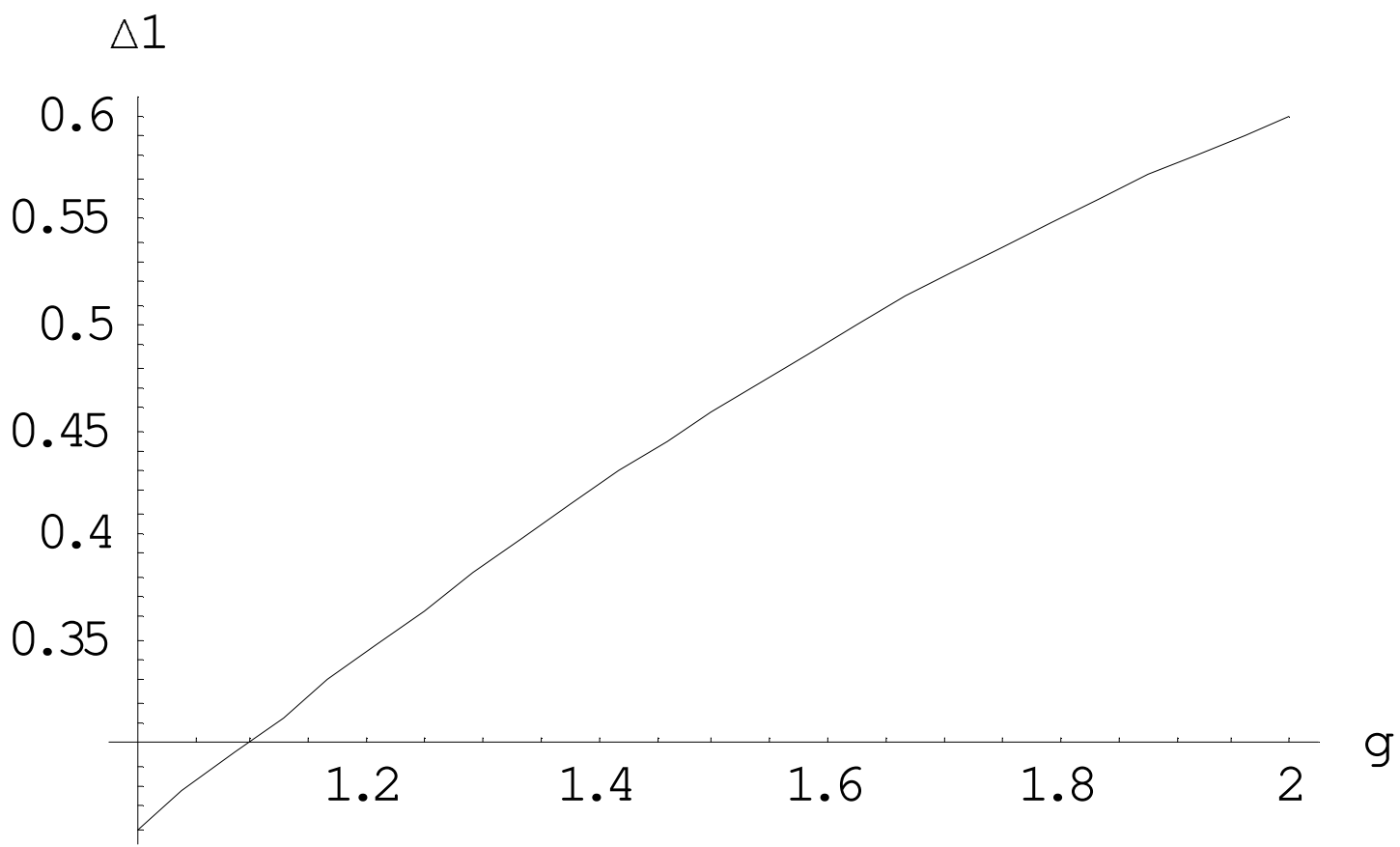
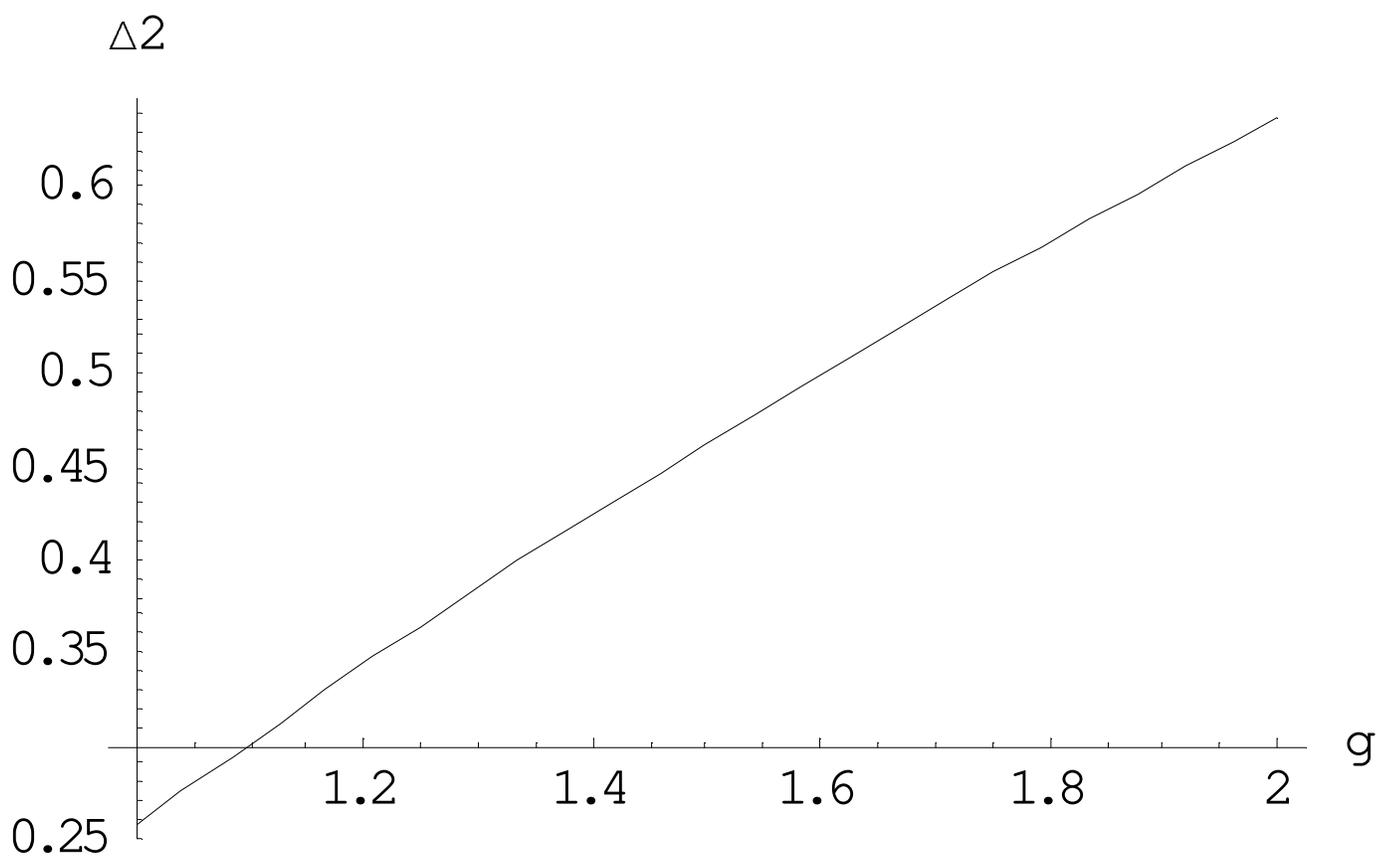
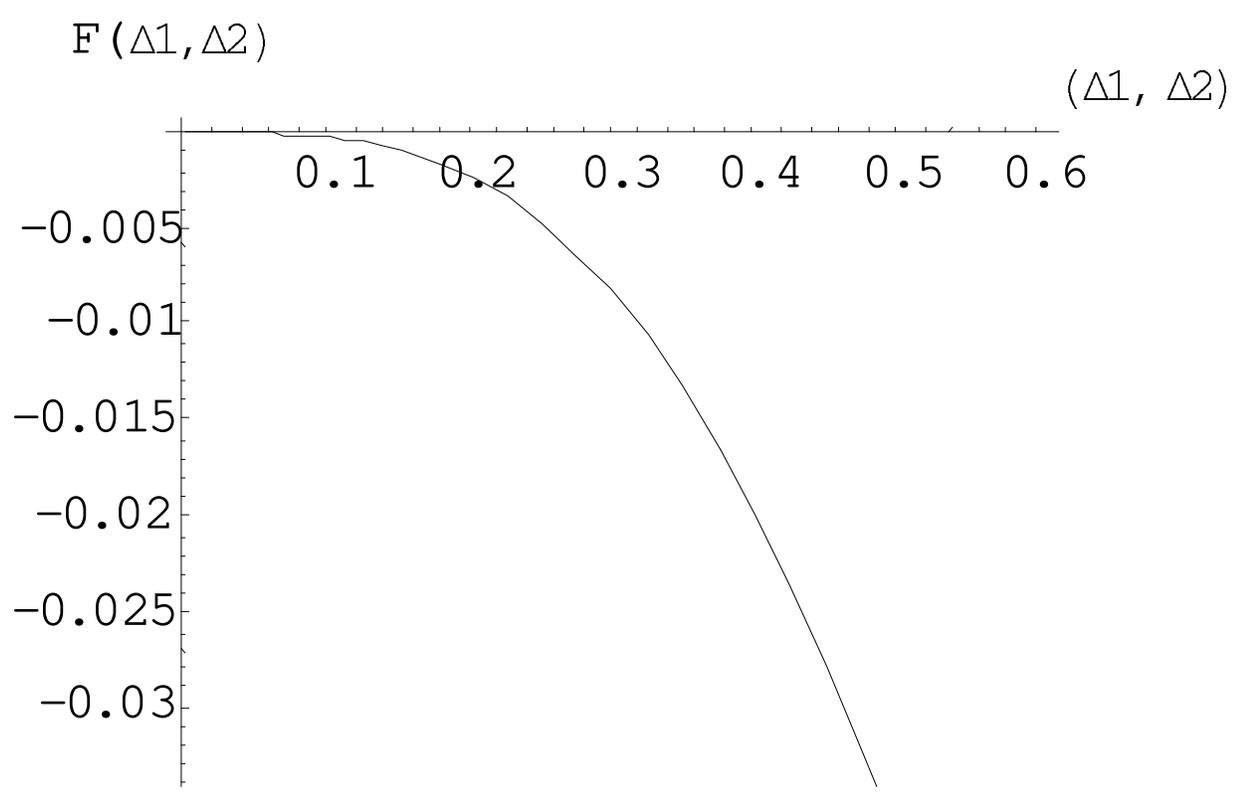
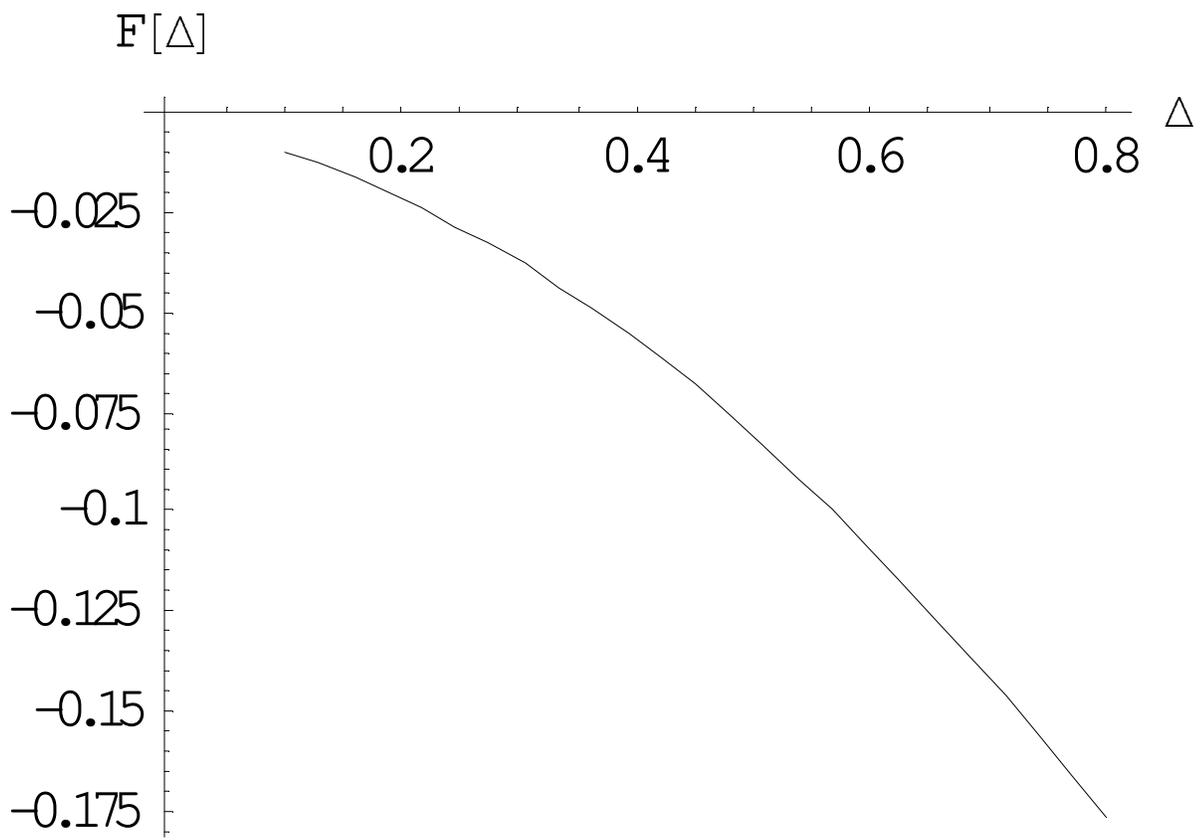
\end{document}